\documentclass[aps,prc,onecolumn]{article}
\usepackage{amsmath}
\usepackage{amssymb}
\usepackage[dvips]{graphicx}
\usepackage{slashed}
\usepackage[utf8]{inputenc}

\newcommand{\vecvar}[1]{\mathbf #1}
\newcommand{\vecp}{\mathbf p}
\newcommand{\vecx}{\mathbf x}
\newcommand{\pint}{\int ~ \frac{d^3 {\mathbf p}}{(2\pi)^3}}

\begin{document}

\title{\bf Shear Viscosity of a Quark Plasma Near the Chiral Phase Transition}
\author{\c{C}a\u{g}lar Do\u{g}an \\
\it Department of Physics, Faculty of Science, Istanbul University, \\
\it Vezneciler 34134 Istanbul, TURKIYE}
\date{}
\maketitle

\begin{abstract}
We calculate the shear viscosity $\eta$ of a quark plasma through the two-flavor Nambu-Jona-Lasinio (NJL) model in the chiral limit, at finite 
temperature $T$, and baryon number chemical potential $\mu_B$. We solve the Boltzmann equation by using in the collision term cross sections that are 
correct to leading order in the coupling constant. We find the ratio of shear viscosity to entropy density above and slightly 
below the chiral transition temperature to vary from 1.5 to 13 times the conjectured lower bound of $(4\pi)^{-1}$ depending on the chemical potential to 
temperature ratio. Ratio of shear viscosity to entropy density is found to be a monotonically increasing function of the ratio of chemical potential to temperature for 
temperatures between 185 and 300 MeV.
\end{abstract}

\section{Introduction}
\label{introduction}

Asymptotic freedom property of Quantum Chromodynamics (QCD) allows rigorous perturbative calculations of transport coefficients to be done at temperatures high compared
to the QCD scale $\Lambda_{QCD}$. Recent work over the last two decades has resulted in the calculation of transport coefficients to leading order in the strong coupling constant for 
high temperature QCD~\cite{leadinglog,leadingorder,bulk,leadinglogchempot}. The drawback of these perturbative methods is that they require the strong coupling constant to 
be perturbatively small. However, at the temperature of 200 MeV attained at the Relativistic Heavy Ion Collider (RHIC) or the higher temperature of 490 MeV that is expected 
to be reached at the Linear Hadron Collider (LHC) this condition is not fulfilled. This motivates the search for alternative methods to calculate these transport 
coefficients at the relevant temperatures. 

One such alternative method is the use of models that possess symmetries of low energy QCD to calculate the transport coefficients of interest. For two flavors of 
massless quarks these symmetries are $\text{SU(2)}_\text{V} \times \text{SU(2)}_\text{A} \times \text{U(1)}_\text{B}$. The proposed model, of course, should also be able 
to account for the spontaneous breaking of the axial symmetries, and the invariance of the ground state under the diagonal subgroup $\text{SU(2)}_\text{V} 
\times \text{U(1)}_\text{B}$. Corresponding to each broken (axial) generator a Goldstone boson appears, which are then naturally identified with the neutral and charged 
pions. 

NJL model~\cite{nambu} which was proposed quite a time ago is one such model. It gives a realistic description of low energy QCD. 
It exhibits a chiral phase transition through which the quarks (and anti-quarks) acquire 
masses as the temperature is lowered below the critical temperature\footnote{Critical temperature will always be used in this article to indicate the 
temperature at which the constituent mass of quarks vanishes.}. Even though it was put forth as a theory of nucleons, it was reinterpreted as a theory of quarks, and 
extended to finite temperature and nonvanishing baryon number chemical potential.

Pions and the sigma meson are not propogating degrees of freedom in this model, rather they appear as poles in the pseudoscalar and scalar channels, respectively, of 
the effective interaction between quarks which is obtained by summing the so-called bubble diagrams. This is the kind of picture that emerges for QCD in the limit of a large 
number of colors $N_c$, so this 
is the limit in which the NJL model is expected to reproduce QCD~\cite{largeN}. This also provides the justification for summing an infinite number of diagrams in the gap 
equation, which contribute at the same order $\mathcal{O}(1)$, to obtain the nonperturbative constituent mass of quarks.

Previous calculations of the shear viscosity of the NJL model include those that were done diagrammatically~\cite{diagramsnjl,diagramsnjl2} both at vanishing and finite 
baryon chemical potentials. In other calculations, even though the transport equation was used, the assumption of relaxation time approximation was 
made~\cite{relaxnjl,shearbulkrelaxnjl,kineticnjl}, and the shear viscosity coefficient was calculated in the so-called random phase approximation to leading order in $N_c$ 
at vanishing chemical 
potential~\cite{relaxnjl2}. There were other calculations at vanishing baryo-chemical potential~\cite{relaxnjlsu3, relaxnjlsu3_2} which considered the enlarged symmetry group of 
$\text{SU(3)}_\text{V} \times \text{SU(3)}_\text{A} \times \text{U(1)}_\text{B}$, and that also assumed the relaxation time approximation to be valid. 

There also exist 
results for the shear viscosity of the quark gluon plasma at leading order of the strong coupling constant~\cite{leadingorder}, albeit at vanishing baryon 
number chemical potential, and at leading logarithmic order at finite baryon number chemical potential~\cite{leadinglogchempot}. These will be compared in 
Section~\ref{results} with the results obtained in this paper as well.

Jeon had shown~\cite{jeonscalar} in the context of $\lambda \phi^4$ theory that the leading order diagrammatic calculation of the viscosity coefficients is 
equivalent to that done with the Boltzmann equation, and later it was shown~\cite{jeonphoto}, at least in the case of photo production, that this holds true for finite 
baryon number chemical potential, too. It will be assumed in this article that the equivalence between the perturbative calculations of the viscosity coefficients 
diagrammatically and via kinetic theory holds even when the baryon number chemical potential is nonzero. 

The aim of this article is twofold: To calculate the shear viscosity of the NJL model at vanishing baryon number chemical potential by solving the 
Boltzmann equation, that is without making the relaxation time approximation, and to systematically generalize it to finite baryon number chemical potentials. In a 
leading order in $N_c$ calculation, summation of bubble diagrams leads to pion and sigma meson exchange in the scalar and pseudoscalar channels, respectively. Above the 
chiral phase transition these appear as resonances, however as the temperature gets closer to the critical temperature the pion resonances turn into bound states that 
may then go on-shell. Although this is innocuous in s-channel diagrams, it leads to divergences in t- and u-channel ones. This problem was not present in previous 
calculations utilizing the relaxation time approximation, because an integrated cross-section was used there. In order to cut off these divergences one needs 
to include the width of the pion and sigma mesons 
below and slightly above the critical temperature. However, this width does not appear at leading order in $N_c$. We choose to avoid these divergences 
instead by approximating the interaction between quarks with the cross section which is correct to leading order in the coupling constant. 

As ensuing analysis of the NJL model~\cite{decoupling} showed, the interaction between quarks and the mesons gets weaker as the temperature approaches the critical 
temperature from below. Therefore, the neglect of bubble diagrams can, at best, be a reasonable approximation as long as the coupling of quarks to the effective mesonic 
degrees of freedom $g_{\pi qq}$ is less than one. This condition will be satisfied if the constituent mass of quarks $M$ is less than the zero-temperature pion decay 
constant $F_\pi \approx 95$ MeV, that is $g_{\pi qq} = M/F_\pi \lesssim 1$. This condition determines the lowest temperature at which we will calculate the shear 
viscosity coefficient in this article.

Dependence of shear viscosity of the NJL model on the temperature will be different depending on 
whether the system is in the chirally symmetric phase or not. Taking into consideration the fact that the 
shear viscosity will be inversely proportional to the square of the coupling constant $G \text{[MeV}^{-2}\text{]}$ of the NJL model, in the phase of broken chiral 
symmetry shear viscosity scales, on dimensional grounds, as,
\begin{eqnarray}
\label{massiveshear}
\eta &=& (G^2 T)^{-1} \times f(\mu /T, M/T) ~ ,
\end{eqnarray}
whereas for the chirally symmetric phase scaling of the shear viscosity is as follows:
\begin{eqnarray}
\label{masslessshear}
\eta &=& (G^2 T)^{-1} \times g(\mu /T) ~ .
\end{eqnarray}

Only two of the three parameters among temperature $T$, quark chemical potential $\mu$, and $M$ in Eq.~(\ref{massiveshear}) are independent. The precise way in which 
the third parameter depends on the others through the gap equation Eq.~(\ref{gapequation}), and the issue of cutoff dependence of the 
results overlooked in this section to simplify matters will be taken up in Sections~\ref{njlmodel} 
and~\ref{overview}, respectively. Section~\ref{njlmodel} will also give a brief description of the NJL model, details of the calculation and how the Boltzmann 
equation is implemented will be explained in Section~\ref{overview}. Section~\ref{results} will be devoted to a discussion of our results.

\begin{figure}
\label{etaovers}
\centering
\includegraphics[width=0.4\textwidth,height=0.2\textheight]{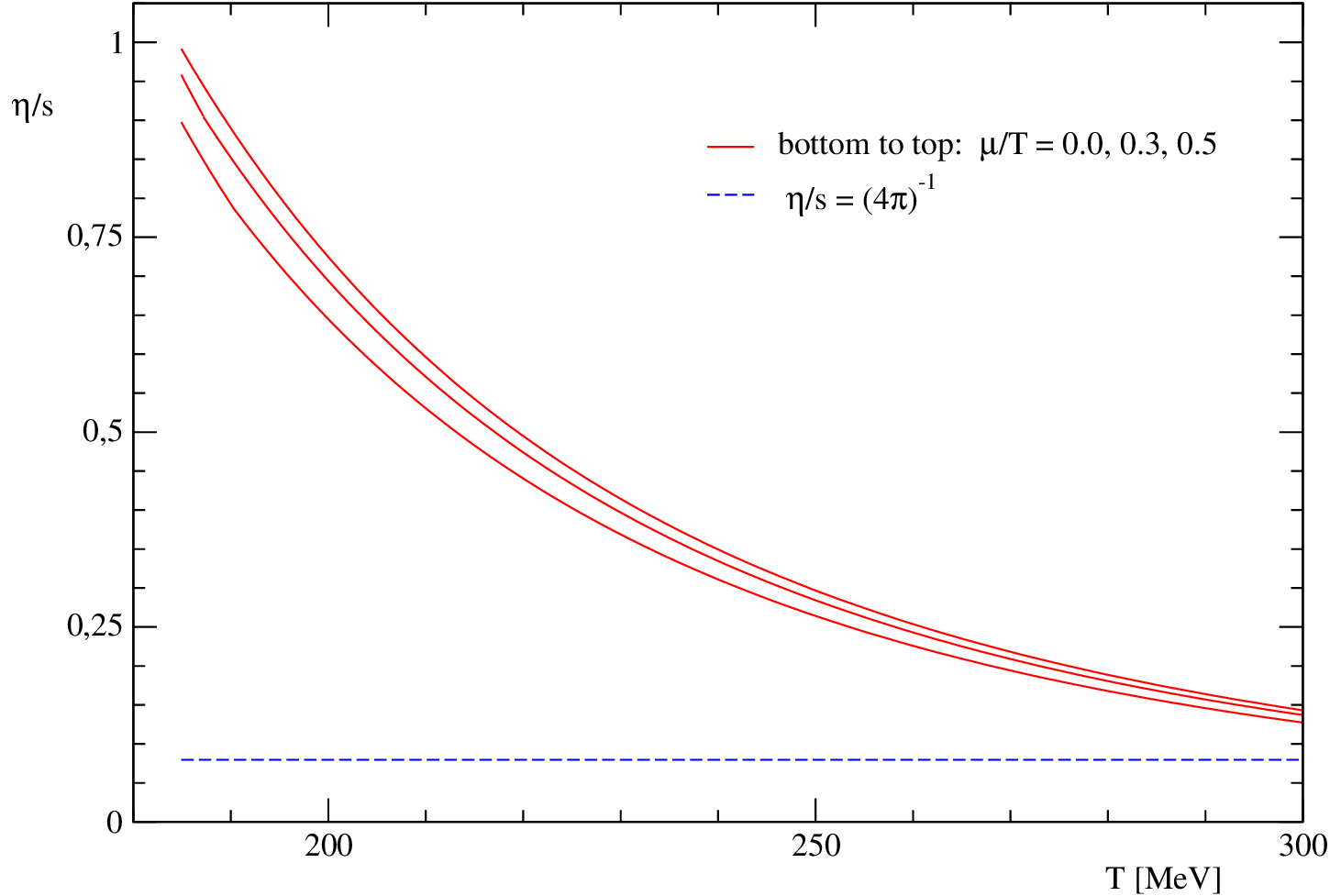}
\includegraphics[width=0.4\textwidth,height=0.2\textheight]{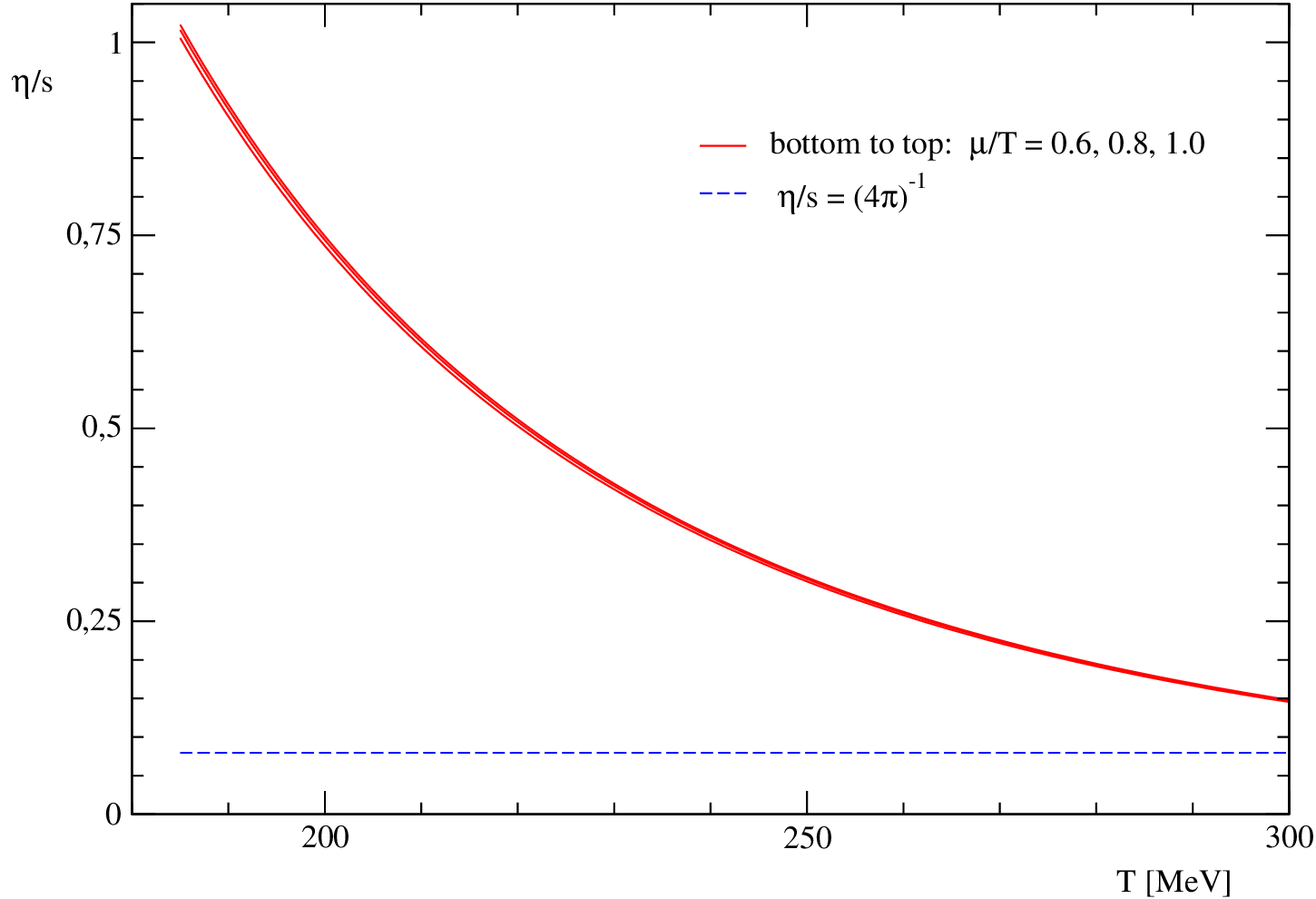}
\caption{Shear viscosity to entropy density as a function of temperature in MeV, shown on the left panel are those for the quark number chemical potential to temperature 
ratios of 0.0, 0.3 and 0.5, whereas on the right panel plotted are those for the ratios of 0.6, 0.8 and 1.0. The horizontal dashed line on each graph indicates the 
conjectured lower bound of $\eta /s = (4\pi)^{-1}$, see references \cite{shearbound}.}
\end{figure}

Our major result in this paper is the calculation of the ratio of the shear viscosity to the entropy density for ratios of chemical potential\footnote{The values quoted
in Fig.~\ref{etaovers} are those of the ratio of quark number chemical potential to temperature.} to temperature ranging from $\mu/T=0.0$ to $\mu/T=1.0$ as shown in 
Fig.~\ref{etaovers}. Although it is hard to tell from the figure, the 
parametric behavior of the ratio of shear viscosity to entropy density $s$ changes at the critical values of $T_c=190$ and $T_c=187$ MeV for $\mu/T=0.0$ and $\mu/T=0.3$, 
respectively. For the higher values of the chemical potential to temperature ratio, the curves shown in the figure correspond to the chirally symmetric phase only. 

\section{Two-Flavor NJL Model}
\label{njlmodel}

NJL model is given by the following Lagrangian density:

\begin{equation}
\label{eq:njlmodel}
{\mathcal L} = {\bar \Psi} \left( i \slashed{\partial} - m \right) \Psi + G \left[ \left( {\bar \Psi} \Psi \right)^2 + \left( {\bar \Psi} i \gamma_5 {\vec \sigma} \Psi 
\right)^2 \right] + {\bar \Psi} \mu \gamma^0 \Psi \ .
\end{equation}

The Lagrangian given above is for $N_f=2$, where $N_f$ is the number of quark flavors. Components of the 
matrix vector denoted by $\vec \sigma$ in the second interaction term above are the  Pauli matrices. Quarks are chosen to be in the fundamental representation of 
SU(3), then the anti-quarks have to be in the conjugate representation. Color indices of the quark field will be suppressed throughout the paper as neither interaction term in the Lagrangian density depends on them.

In the chiral limit, the mass matrix is zero $m=0$. The last term has to be introduced when the chemical potential corresponding 
to baryon number is nonzero. In order to preserve the symmetries of the model it will be chosen to be proportional to the identity matrix (in flavor space), with the 
proportionality constant $\mu$ being the common chemical potential of both flavors. The relation between the chemical potential of the quark species and that for baryon 
number chemical potential $\mu_B$ is $\mu = \mu_B /3$.

NJL model in the chiral limit has two phases in the $\mu-T$ plane separated by a second order phase transition\footnote{The nature of the transition is very sensitive to the 
value of the current quark mass, and changes from a second order transition to either a first order transition or a cross-over for finite values of this parameter depending 
on the temperature to chemical potential ratio.}. The existence of 
a nontrivial solution to the gap equation given below determines whether the system is in the chirally symmetric high-temperature phase or the low-temperature phase of 
broken chiral symmetry\footnote{The coefficient in the gap equation should actually be $N_c N_f + \frac{1}{2}$, however the second term is subleading for large $N_c$ and is 
generally dropped.}.

\begin{eqnarray}
\label{gapequation}
M \left[ 1 - 4G N_c N_f \int_0^\Lambda \ \frac{d^3 \vecp}{(2\pi)^3 \ E_p} \left[ 1-n_F(T,\mu)-{\bar n}_F(T,\mu) \right] \right] = 0
\end{eqnarray}
Here $E_p=(p^2+M^2)^{1/2}$ is the quasiparticle energy and $p \equiv |\vecp|$ denotes the magnitude of the
particle's momentum, $n_F(T,\mu)=\left[e^{(E_p-\mu)/T}+1\right]^{-1}$ is the distribution function of quarks, and ${\bar n}_F(T,\mu)=n_F(T,-\mu)$ is that of anti-quarks in 
thermodynamic equilibrium. Eq.~(\ref{gapequation}) always has the trivial solution $M=0$ and for a given chemical potential this is the only solution above the critical 
temperature. However, at fixed chemical potential, as the system cools down the ground state that minimizes the free energy will lead to a nonzero value of the 
constituent quark mass, i.e. spontaneous chiral symmetry breaking.

As described in greater detail elsewhere~\cite{klevansky}, an analysis of the NJL model in the chiral limit at vanishing temperature and chemical potential determines the 
value of the coupling constant G and the 3-momentum cutoff $\Lambda$ in Eq.~(\ref{gapequation}) as $G \Lambda^2 = 2.14$ and $\Lambda = 653$ MeV, respectively. Therefore, 
this effective theory can be used for the calculation of shear viscosity for energies below this cutoff scale.

\section{Details of the Calculation}
\label{overview}

The cross section for a given particle to scatter off other particles in the 
plasma scales as\footnote{The estimate given above assumes that $T \ll \Lambda$, otherwise the scattering cross section will also depend on the cutoff.} 
$\sigma \sim G^2 T^2 \sim (GT^2)^2 T^{-2}$. This cross section 
then leads to a mean free path of $l \sim (T^3 (GT^2)^2 T^{-2})^{-1} \sim (GT^2)^{-2} T^{-1}$. Since this mean free path is much larger than the thermal wavelength of 
quasi-particles in the plasma $l \sim (GT^2)^{-2} T^{-1} \gg T^{-1}$,  
one is justified in using kinetic theory\footnote{In a leading order in $N_c$ calculation with $N_c \gg 1$, but $G \Lambda^2 N_c \sim G T^2 N_c \sim \mathcal{O}(1)$ these 
estimates would have been 
$\sigma \sim (GT^2N_c)^2 T^{-2} N_c^{-1} \sim \mathcal{O}(N_c^{-1})$, $l \sim (GT^2N_c)^{-2} T^{-1} N_c \sim \mathcal{O}(N_c) \gg T^{-1}$, and 
$\eta \sim (GT^2N_c)^{-2} T^3 \sim \mathcal{O}(1)$.}, and more specifically the Boltzmann equation to calculate transport coefficients. The above estimate for the mean free 
path then leads to the following estimate for shear viscosity: $\eta \sim l ~\mathcal{E} \sim (GT^2)^{-2} T^{-1} T^4 \sim (G^2 T)^{-1}$, where $\mathcal{E}$ is the energy 
density of the quark plasma. This estimate was the basis for Eqs.~(\ref{massiveshear}) and (\ref{masslessshear}).

We denote the phase space density of both quarks and anti-quarks in the thermal bath by 
$f(\vecx,\vecp)$. In the hydrodynamic limit of small deviations from local thermodyanmic equilibrium one approximates 
this phase space density as a local equilibrium piece $f_{eq}(\vecx,\vecp)$ and a small deviation from that $\delta f(\vecx,\vecp)$. The local equilibrium distribution, of 
course, has to be taken equal to $n_F(T,\mu)$ and $\bar{n}_F(T,\mu)$ for quarks and anti-quarks, respectively. For a 
divergenceless flow relevant for shear viscosity this phase space density can be taken to be 
independent of time.

\begin{eqnarray}
f(\vecx,\vecp) &=& f_{eq}(\vecx,\vecp) + \delta f(\vecx,\vecp)
\end{eqnarray}
The Boltzmann equation gives the change in the phase space density as a result of collisions, which 
are encapsulated by the collision term $C[f]$ as follows:
\begin{eqnarray}
\label{boltzmanneq}
\frac{\partial f}{\partial t} + \vecvar{v}_\vecp \cdot {\boldsymbol \nabla}_\vecx f = -C[f] ~.
\end{eqnarray}
The velocity of a particle with momentum $\vecp$ in the above equation is $\vecvar{v}_\vecp = 
\vecp/E_p$ as usual for a relativistic particle. As the LHS of Eq.~(\ref{boltzmanneq}) is already first 
order in the gradients of the flow velocity, the local equilibrium distribution function should be 
used there for a leading order calculation. We use the invariance of the Boltzmann equation under 
boosts (more generally Lorentz 
transformations) to evaluate the derivatives on the LHS in a reference frame where the flow velocity 
is nonzero, and specialize to the local rest frame only after this has been done. This gives

\begin{eqnarray}
\vecvar{v}_\vecp \cdot {\boldsymbol \nabla}_\vecx f_{eq}(\vecx,\vecp) = \beta f_{eq} 
(1-f_{eq}) \frac{|\vecp|^2}{E_p} \left[ \frac{1}{\sqrt 6} \left( {\nabla}_i u_j + 
{\nabla}_j u_i - \frac{2}{3} \delta_{ij} {\boldsymbol \nabla} \cdot \vecvar{u} \right) \right] I_{ij}
\end{eqnarray}
with the second rank tensor (under rotations) $I_{ij}$ in the above equation defined as
\begin{eqnarray}
I_{ij}(\vecp) \equiv \sqrt{\frac{3}{2}} \left({\hat p}_i {\hat p}_j -\frac{1}{3} 
\delta_{ij}\right)
\end{eqnarray}
Normalization is chosen such that $I_{ij}(\vecp) I_{ij}(\vecvar{k})=
P_2(cos\theta_{\vecp \vecvar{k}})$, where $P_2(cos\theta)$ is the second Legendre polynomial. We follow the conventions of~\cite{leadingorder} in the definition and 
normalization of the above tensor. The 
departure from equilibrium at linearized order can naturally be parametrized as follows:  
\begin{eqnarray}
\delta f^a(\vecp) = \beta^2 f_{eq}^a(p) \left[1-f_{eq}^a(p)\right]
\left[\frac{1}{\sqrt 6} \left( {\nabla}_i u_j+{\nabla}_j u_i - \frac{2}{3} \delta_{ij} 
{\boldsymbol \nabla} \cdot \vecvar{u} \right) \right] \chi^a_{ij}
\end{eqnarray}
This leads to the simpler equation given below. The expression for the linearized collision operator 
$C$ appearing in the equation below will be given later.  
\begin{eqnarray}
\label{boltzmannreduced}
S^a_{ij}(\vecp) = (C \chi_{ij})^a(\vecp)
\end{eqnarray}
where the source term $S_{ij}^a$ is defined to be
\begin{eqnarray}
S_{ij}^a = -T \frac{|\vecp|^2}{E_p} f_{eq}^a(p) \left[1-f_{eq}^a(p)\right] I_{ij}(\vecp)
\end{eqnarray}
and due to the rotational invariance of the collision operator, once the tensorial structure of 
$\chi_{ij}$ is peeled off it can only be a function of the magnitude of momentum in the local rest 
frame.
\begin{eqnarray}
\chi^a_{ij} = I_{ij}(\vecp) \chi^a(p)
\end{eqnarray}
It's convenient to define the following inner product
\begin{eqnarray}
\left( \phi_{ij}, S_{ij}\right) \equiv -\beta^2 \sum_a \pint f_{eq}(p) \left[1-f_{eq}(p)\right] 
\frac{|\vecp|^2}{E_p} \phi(p) 
\end{eqnarray}
where we used $I_{ij}(\vecp) I_{ij}(\vecp)=1$. Expanding the solution in terms of basis functions 
given below
\begin{eqnarray}
\label{expansion}
\chi^a(p) = \sum_{m=1}^K {\tilde \chi}^a_m \phi_{(m)}(p)
\end{eqnarray}
gives the vector representation of the source and the matrix representation of the linearized 
collision operator in this basis as follows.
\begin{eqnarray}
\nonumber (\chi_{ij},S_{ij}) &=& \sum_{a,m} {\tilde \chi}^a_m {\tilde S}^a_m \\
(\chi_{ij},C \chi_{ij}) &=& \sum_{a,m} \sum_{b,n} {\tilde \chi}^a_m {\tilde C}^{ab}_{mn} 
{\tilde \chi}^b_n
\end{eqnarray}
In Eq.~(\ref{expansion}), index $a$ denotes the type of particle, in our case either quark or 
anti-quark, for which this is the deviation from equilibrium and the lower index stands for the 
coefficient of the basis function $m$ in the expansion of the deviation from equilibrium. Here is the explicit expression that we postponed to give for the linearized collision 
operator ${\tilde C}_{mn}^{ab}$ appearing above.
\begin{eqnarray}
\label{collisionmatrix}
\nonumber {\tilde C}_{mn}^{ab} &=& \frac{1}{8} \sum_{a,b,c,d} 
\int_{\vecp ,\vecvar{k},\vecvar{p^\prime},\vecvar{k^\prime}} f_{eq}^a(E_p) f_{eq}^b(E_k) 
\left[1-f_{eq}^c(E_{p\prime})\right] \left[1-f_{eq}^d(E_{k\prime})\right] 
\left|\mathcal{M}^{ab}_{cd}\right|^2 \\ 
\nonumber && \times \left[\phi_{(m)}^a I_{ij}(\vecp)+\phi_{(m)}^b I_{ij}(\vecvar{k})-
\phi_{(m)}^c I_{ij}(\vecvar{p^\prime})-\phi_{(m)}^d I_{ij}(\vecvar{k^\prime})\right] \\
\nonumber && \cdot \left[\phi_{(n)}^a I_{ij}(\vecp)+\phi_{(n)}^b I_{ij}(\vecvar{k})-
\phi_{(n)}^c I_{ij}(\vecvar{p^\prime})-\phi_{(n)}^d I_{ij}(\vecvar{k^\prime})\right] \\
&& \times (2\pi)^4 \delta^{(4)}\left(P+K-P^\prime-K^\prime\right)
\end{eqnarray}
The above integrals can be simplified considerably, details of the simplified collision term are given in Appendix~\ref{integrationvariables}. One basis that has 
the right asymptotic behavior and converges with a sufficiently small number of basis elements is 
provided by the following functions.
\begin{eqnarray}
\phi_{(m)}(p) = \frac{p\left(p/T\right)^{m-1}}{\left(1+p/T\right)^{K-2}}
\end{eqnarray}
with $m=1,2,...,K$. In our case, when the cutoff is neglected choosing five basis elements, 
that is $K=5$, leads to one part in a thousand accuracy, whereas inclusion of the cutoff severely 
reduces this to slightly above $1.5\% $. Shear viscosity is then given in terms of the source vector 
and the inverse of the collision matrix in this basis as follows:
\begin{eqnarray}
\eta = \frac{1}{15} {\tilde \chi}^T {\tilde S} = \frac{1}{15} {\tilde S} {\tilde C}^{-1} {\tilde S} 
\end{eqnarray}

A few comments are in order about the matrix elements used in Eq.~(\ref{collisionmatrix}), in the 
collision term of the Boltzmann equation. Since we have worked to leading order in $N_c$ and treated 
$G\Lambda^2 N_c \sim \mathcal{O}(1)$ in the gap equation, consistency would require that we include in the 
matrix elements all scattering diagrams to leading order in $N_c$, that is the so-called bubble 
diagrams, for a given process as well. These diagrams, when summed, 
would then lead to the effective interaction of a pion being exchanged in the pseudo-scalar channel, 
and a sigma meson being exchanged in the scalar channel. Although this is the correct procedure to follow, we will ignore the pion 
(or the sigma meson) pole contribution, and approximate the quark and anti-quark interactions by the tree level term read off from the NJL model lagrangian. We 
are going to compare our results for the zero baryon chemical potential case to previous work in Section~\ref{results} to see how this approximation affects our results.

It is in principle possible to include the pion pole contribution to the scattering matrix elements in 
the calculation of the shear viscosity of the NJL model, however it should be noted that this is 
nontrivial due to the divergences it causes in the collision term of the Boltzmann equation. We work 
in the chiral limit throughout the paper, and in this limit the pion is exactly massless. Therefore, 
the divergences in the s-channel pion exchange is harmless due to shrinking of the phase space, and
a bare pion propogator is sufficient for s-channel exchange. This parallels s-channel gluon exchange 
in a quark gluon plasma calculation of shear viscosity where it suffices to use a bare gluon propogator. 
However, t- and u-channel pion exchange lead to divergences unless the imaginary part of the 
pion propogator, that is generated by scatterings of the pion off quarks and anti-quarks in the 
plasma, is included in those matrix elements. This is expected to arise at order $\mathcal{O}(N_c^{-2})$ for the cross-sections, that is at next-to-leading order in $N_c$. 

As for the masses of the species in the matrix elements and in the equilibrium Fermi-Dirac 
distribution functions, these are taken to be zero in the chirally symmetric phase, and non-zero and 
equal to the value obtained from Eq.~(\ref{gapequation}) for a given temperature and chemical potential 
in the calculation of shear viscosity explained in this paper. For our parameter 
choice, the critical temperatures for chiral symmetry breaking at $\mu/T = 0.0$ and $\mu/T = 0.3$ are 
$T_c=190$ and $T_c=187$ MeV, respectively. The critical temperature is below 185 MeV for all the higher 
ratios of the chemical potential to temperature ratio, so the quasi-particles are taken to be massless 
in those cases. 

It should be emphasized that the masses of the quasiparticles in the NJL model are nonperturbative masses, 
and thus they are included in the low temperature phase where they are nonzero. The treatment of masses in 
this model should be 
contrasted with how the thermal masses of species were handled in a leading order in 
the coupling constant calculation of the shear viscosity of the quark gluon plasma 
$m_{th}^2 \sim g_s^2 T^2$ and of scalar theory with quartic self-couplings 
$m_{th}^2 \sim \lambda T^2$, where $g_s$ and $\lambda$ are the coupling constants of QCD and 
$\lambda \phi^4$ theory, respectively. In those calculations, all on-shell particles were taken 
to be exactly massless, and the perturbative effects induced by the thermal masses rightly 
belonged among the corrections higher order in the coupling constant the leading order result receives.

\begin{table}[t]
\begin{center}
\begin{tabular}{|c|c|c|}
\hline & ${\mathcal M}^{ab}_{cd}$ & $|{\mathcal M}^{ab}_{cd}|^2/G^2$ \\
\hline $qq \leftrightarrow qq$ & ${\mathcal M}_{qq}(P,K;P^\prime,K^\prime)$ & 
$256 N_c^2 \left(t^2+u^2-tu+6M^4\right)$ \\
\hline ${\bar q} {\bar q} \leftrightarrow {\bar q} {\bar q}$ & 
${\mathcal M}_{{\bar q}{\bar q}}(P,K;P^\prime,K^\prime)={\mathcal M}_{qq}(P^\prime,K^\prime;P,K)$ & 
$256 N_c^2 \left(t^2+u^2-tu+6M^4\right)$ \\
\hline $q {\bar q} \leftrightarrow q {\bar q}$ & 
${\mathcal M}_{q{\bar q}}(P,K;P^\prime,K^\prime)={\mathcal M}_{qq}(P,-K^\prime;P^\prime,-K)$ & $256 N_c^2 \left(t^2+s^2-ts+6M^4\right)$ \\
\hline
\end{tabular}
\end{center}
\label{crosssections}
\caption{Charge conjugation invariance forces the matrix elements for scattering of quarks off quarks and of antiquarks off antiquarks to be the same, whereas crossing 
symmetry relates the matrix elements for scattering of quarks off quarks to scattering of quarks off antiquarks. The last column 
gives the absolute square of the matrix elements divided by the square of the coupling constant for these processes. The square of the matrix elements are 
summed, but not averaged, over the spins, colors and flavors of the participants. In particular, the matrix elements squred scale wih $N_c^2$ because an average over the 
colors has not been taken. Of course, the number of colors will be set equal to 3.}
\end{table}

Charge conjugation invariance of the kinetic and interaction terms in the NJL model, that is barring the chemical potential term, implies that the matrix elements for the 
scattering of quarks off quarks and of anti-quarks off anti-quarks are the same. Moreover, the matrix 
elements for the scattering of quarks off anti-quarks is related to the above by crossing symmetry. 
Explicit expressions, in terms of the Mandelstam variables, for the squares of these matrix elements as they 
appear in the Boltzmann equation, that is summed but not averaged over the spins, flavors, and colors of the 
participants, are given in the Table 1.

\section{Results \& Discussion}
\label{results}

\begin{figure}
\centering
\includegraphics[width=0.4\textwidth,height=0.2\textheight]{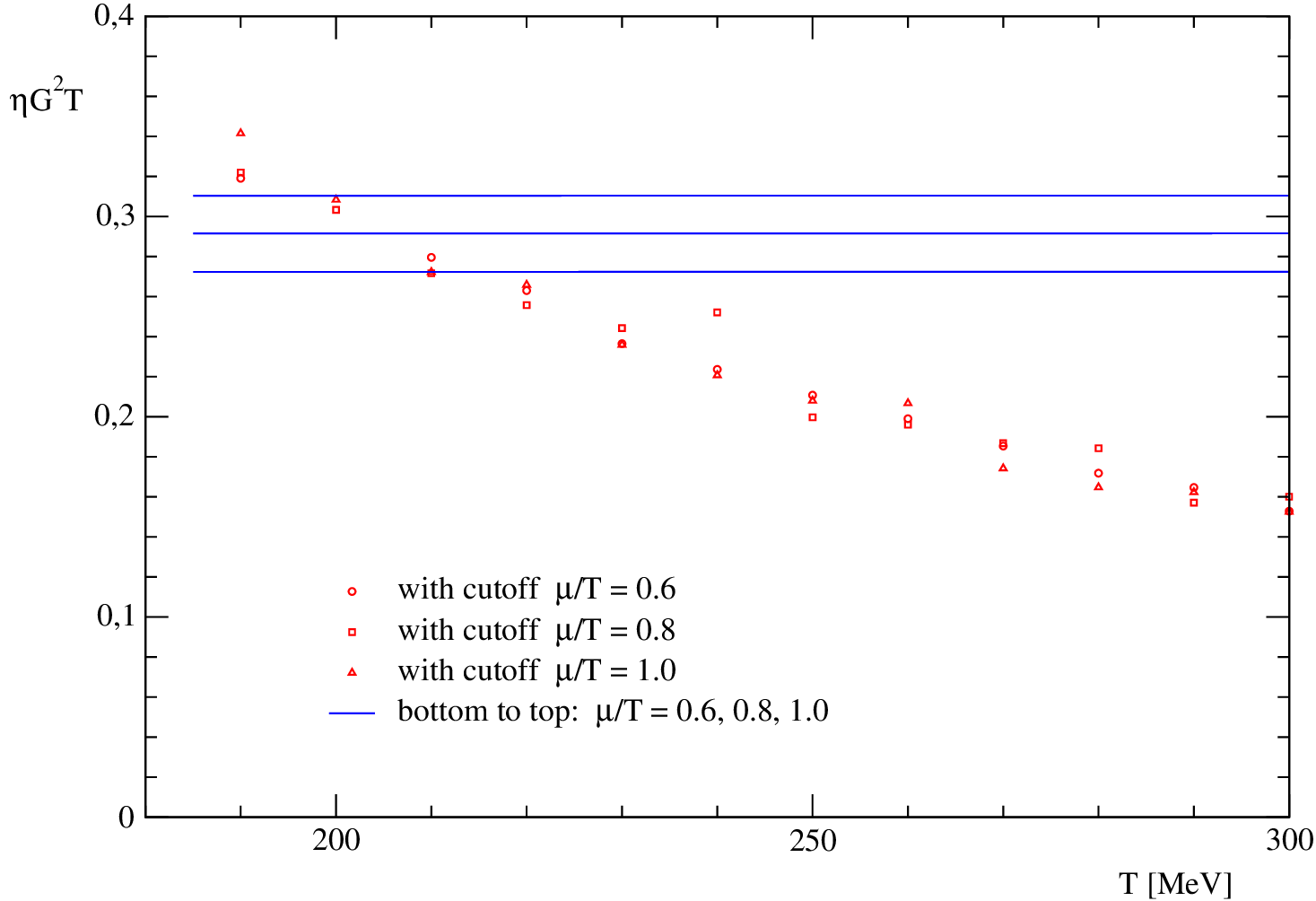}
\caption{The dimensionless product of shear viscosity, coupling constant squared and the temperature 
as a function of temperature in MeV. The horizontal lines in the plot show the constant values of 
this product when the cutoff is taken to infinity for the chemical potential to temperature ratios 
of 0.6, 0.8 and 1.0. The scattered points, on the other hand, are data for the same 
ratios of the chemical potential to temperature when a momentum cutoff is used in the Boltzmann equation.}
\label{fig:shearGsquaretemp}
\end{figure}

In order to understand the manner ratio of shear viscosity to temperature cubed depends on temperature, 
we begin by analyzing the dimensionless product $\eta G^2 T$, this product is expected to only depend 
on the ratio of chemical potential to temperature in the chirally symmetric phase, and should therefore 
be constant for each value of 
this ratio. This is clearly the case when the 3-momentum cutoff is taken to infinity in the Boltzmann equation as can be seen in Figure~\ref{fig:shearGsquaretemp} for 
various values of the chemical potential to temperature ratio.  Thus, the ratio of shear viscosity to temperature cubed scales as $\eta /T^3 \sim T^{-4}$ in the high 
temperature phase. The reason the ratio of shear viscosity to the cube of the temperature varies with temperature at all in the massless phase is because the effective 
coupling constant $GT^2=G\Lambda^2(T/\Lambda)^2=2.14(T/\Lambda)^2$, which is what is relevant for scattering cross-sections, increases with increasing temperature. 

\begin{figure}
\centering
\includegraphics[width=0.4\textwidth,height=0.2\textheight]{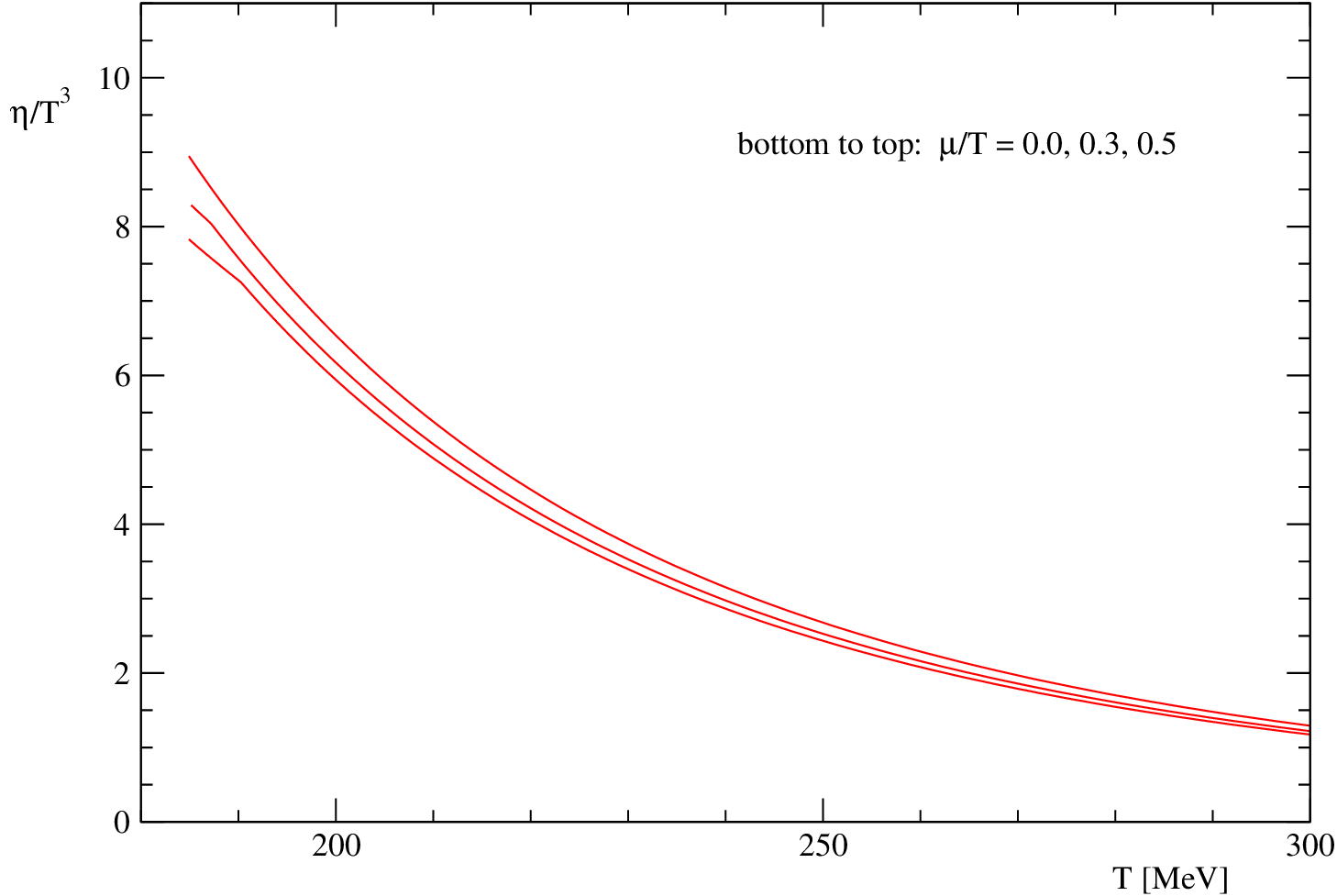}
\includegraphics[width=0.4\textwidth,height=0.2\textheight]{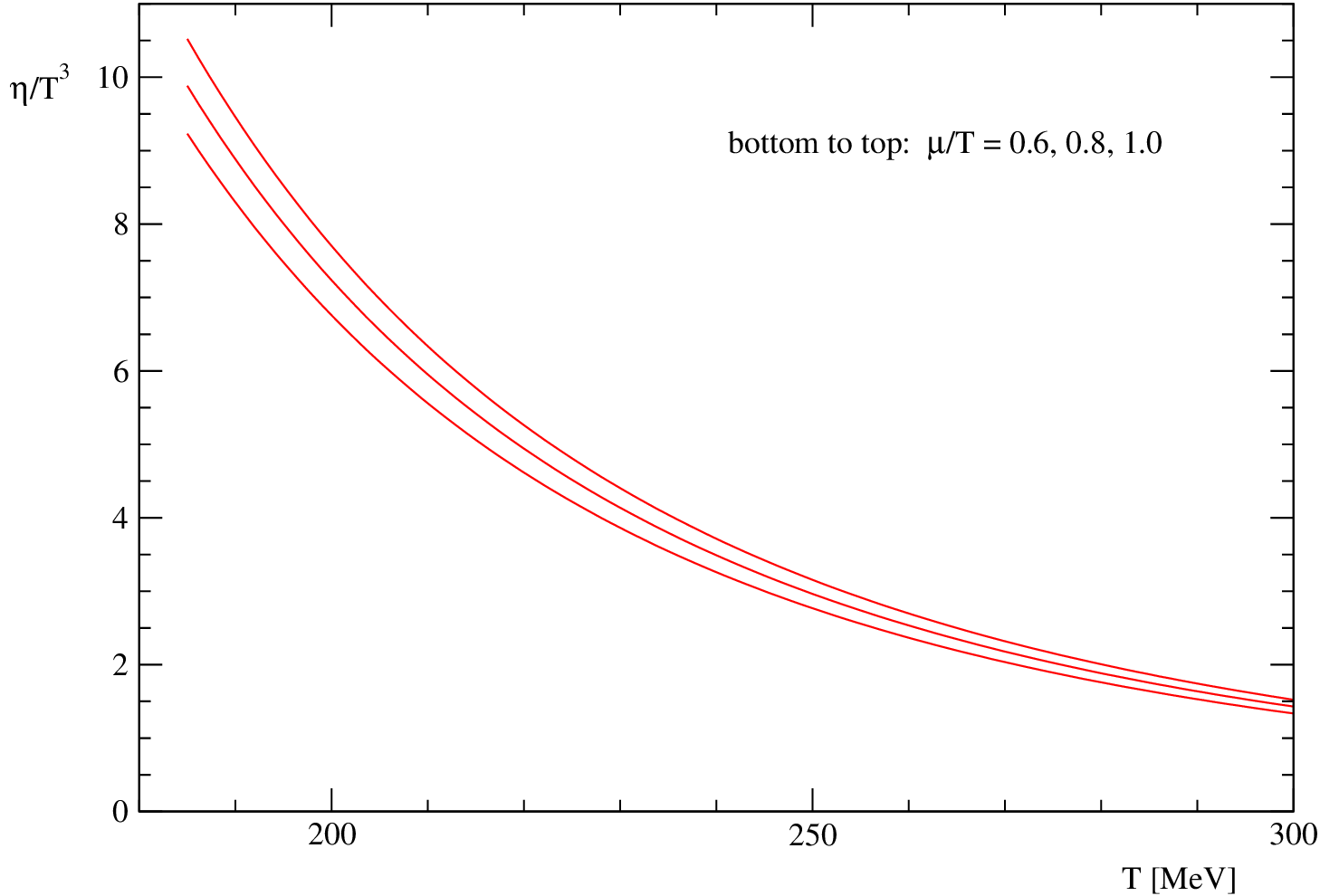}
\caption{Ratio of shear viscosity to temperature cubed as a function of temperature in MeV. The plot on the left 
panel shows the curves for the quark chemical potential to temperature ratios of 0.0, 0.3 and 0.5, whereas on the right panel displayed are those for 0.6, 0.8 and 1.0. Note 
the discontinuity near 190 MeV in the slope of the lower two curves shown on the left panel.}
\label{fig:shearovertempcube}
\end{figure}

We evaluated the shear viscosity coefficient in the phase of broken chiral symmetry only for the chemical potential to temperature ratios of 0.0 and 0.3, as the 
transition temperature in all the other cases falls below 185 MeV. This is the temperature for the vanishing chemical potential case at which the pion-quark coupling 
constant $g_{\pi qq}$ given by the ratio of 
the constituent mass of quarks to the zero temperature pion decay constant $F_\pi$ is of order one $g_{\pi qq}=M/F_\pi \sim \mathcal{O}(1)$, and the "fine structure 
constant`` is negligible $\alpha \equiv g^2_{\pi qq}/(4\pi)$. Our calculation can, at best, be expected to be a good approximation down to these temperatures. Since this 
temperature is very close to the transition temperatures of 190 and 187 MeV for the vanishing chemical potential and for the chemical potential to temperature ratio of 
0.3, respectively, the dimensionless product of the shear viscosity, coupling constant squared and the temperature $\eta G^2 T$ can be Taylor expanded around its value in 
the chirally symmetric phase $\eta G^2 T_{T<T_c} \approx \eta G^2 T_c - a \cdot (T_c-T)$ with a positive constant $a>0$. We found that the values of this dimensionless 
product were fitted very well with a line in the low temperature phase, the slope of this best fit is the above constant $a$. Ratio of shear viscosity to the cube of the 
temperature, unlike the dimensionless product, increases as the temperature is lowered in the massive phase. However, the scaling with temperature changes to 
$\frac{\eta G^2}{T^3} \approx \frac{\left( \eta G^2 -a \right)T_c}{T^4} + \frac{a}{T^3}$. The change in the scaling of the ratio of shear viscosity to temperature cubed in 
the two phases is apparent as there is a discontinuity in the slope of the lowest two curves in the left panel of Fig.~\ref{fig:shearovertempcube} at the transition points 
of 187 and 190 MeV. These discontinuities persist in the lowest two curves in the left panel of Fig.~\ref{etaovers}, too, though they are not as prominent.

Shear viscosity of the NJL model at vanishing chemical potential was calculated previously by Zhuang et 
al.~\cite{relaxnjl2}. What we found for the ratio of shear viscosity to temperature cubed in the low temperature 
phase of broken symmetry agrees with their result exactly. Although they use the same matrix elements in the massive phase as we do, this could not be expected a priori 
given the relaxation time approximation they made in their calculation. On the other hand, the dependence of this ratio on the temperature is quite 
different for the two calculations in the chirally symmetric high temperature phase. Shear viscosity 
scaled with the third power of the temperature decreases with increasing temperature in both 
calculations, yet the scalings are different. In this paper, we find this ratio to decrease with the 
fourth power of the temperature $\eta/T^3 \sim T^{-4}$, whereas the figure in the paper of Zhuang et 
al. indicates a less steep decrease with increasing temperature than ours. Thus, our 
calculation underestimates this ratio by a factor of nearly 3.5 compared to theirs at the temperature 
of 300 MeV.

Notwithstanding the fact that the ratio of shear viscosity to entropy density was calculated previously, it should be emphasized that this ratio calculated with the NJL 
model accounts for the almost saturation of the lower bound of this value observed in experiments conducted at RHIC much better than perturbative QCD calculations do. For 
example, the leading order in the strong coupling constant calculation of Arnold et al.~\cite{leadingorder} found the viscosity coefficient to be 
$\eta = \frac{T^3}{g^4} ~ f(m_D/T)$ where $m_D$ is the Debye mass. The value of the ratio of Debye mass to temperature for two 
flavors considered in this paper is approximately\footnote{$m_D^2=\frac{4\pi}{3}(3+N_f/2)\alpha_s T^2$} $m_D/T \approx 2.24$ for the phenomenological value of 
$\alpha_s \simeq 0.3$. Figure 2 in the aforementioned reference gives for the value of this function $f(m_D/T \approx 2.24) \approx 150$ so that the ratio of shear viscosity 
to entropy density comes out to be $\eta /s \approx 2.60$ which is nearly 3 times the highest value of this ratio obtained in this paper for the vanishing chemical 
potential case. On the other hand, the lowest value of the shear viscosity to entropy density is attained at the upper end of the temperature range, and is $\eta /s \approx 
0.13$ which is only 60$\% $ greater than the conjectured lower bound~\cite{shearbound}. The scalings of the ratio of shear viscosity to the cube of the temperature is also 
different in the two calculations, high temperature perturbation theory predicts this ratio to be independent of temperature, save for the logarithmic running of the 
coupling constant with temperature, whereas in this article we found this ratio to decrease with the fourth power of the temperature $\eta/T^3 \sim T^{-4}$. Finally, the 
ratio of shear viscosity to entropy density increases monotonically as the ratio of chemical potential to temperature is increased from $\mu/T=0.0$ to $\mu/T=1.0$.

If the temperatures at which we evaluated the shear viscosity coefficient had been much smaller than the 3-momentum cutoff $T \ll \Lambda$ that appears in the gap equation
Eq.(\ref{gapequation}), and which defines the theory, the precise value of the cutoff would have been immaterial. This is because the equilibrium distribution 
functions in the collision term of the Boltzmann equation decay exponentially for energies much larger than the temperature $E_p \gg T$ and particles with such high momenta 
do not contribute to the integrals. However, for the temperatures of 185 to 300 MeV at which we are interested in calculating the shear viscosity coefficient 
one has to assess how sensitive one's results are to the value of this cutoff. This is exactly what the scattered points in 
Figure~\ref{fig:shearGsquaretemp} indicate. As can be seen from the figure, at lower temperatures imposing a 3-momentum cutoff tends to increase the dimensionless product, 
and thus the viscosity coefficient itself, whereas at higher temperatures it tends to reduce the value of the shear viscosity to almost half the value it has without the 
3-momentum cutoff.

In this article we calculated the leading order in the coupling constant shear viscosity coefficient of a plasma of quarks and anti-quarks interacting according to the NJL 
model at finite temperature above and slightly below the chiral phase transition temperature and at nonzero 
baryon number chemical potential. In the calculation the ratio of chemical potential to temperature was not greater than one, i.e. $\mu/T \leq 1$. We did not sum all 
scattering diagrams to leading order in the number of colors $\mathcal{O}(N_c^{-1})$, and instead evaluated the matrix elements for scattering by the tree level scalar 
and pseudoscalar interaction terms in the Lagrangian, that is to leading order in the coupling constant $\mathcal{O}(G^2)$. In order to obtain the 
shear viscosity coefficient we solved the linearized Boltzmann equation. We found that the ratio of shear viscosity to entropy density decreased with increasing temperature, 
scaling like $\eta /s \sim T^{-4}$ in the high temperature phase. The value of this ratio was found to be roughly 1.5 times the conjectured lower bound of $(4\pi)^{-1}$ at 
the temperature of 300 MeV for the vanishing chemical potential case, and was not in any of the cases considered greater than 13 times this bound. Therefore, shear viscosity 
over entropy density calculated in this model reproduced the low value observed in experiments far better than perturbative calculations did. Shear viscosity did not change 
much as the ratio of chemical potential to temperature was varied, however it was a monotonically increasing function 
of this variable. 

The vanishing width of the pion as the critical temperature is approached from above prevented us from using all cross sections for scattering that are correct to order 
$\mathcal{O}(N_c^{-1})$ in the Boltzmann equation. As even stable particles acquire a width in a thermal medium, it might be possible to correctly incorporate this width of 
the pion in the calculation of the shear viscosity by including diagrams of $\mathcal{O}(N_c^{-2})$. This nontrivial extension of the current calculation of shear viscosity 
is left for future work.

\appendix
\section{Integration Variables and Limits}
\label{integrationvariables}

In order to simplify Eq.~(\ref{collisionmatrix}) we follow the steps outlined in the appendix of Arnold, et al.~\cite{leadingorder}, 
but generalize their treatment to take into account the finite mass of quarks. For terms that 
are proportional to $t=(P-P^\prime)^2$ the spatial delta function can be used to perform the 
${\mathbf k^\prime}$ integration and the ${\mathbf p^\prime}$ integration can then be shifted into 
one over ${\mathbf q}={\mathbf p^\prime}-{\mathbf p}$. This reduces Eq.~(\ref{collisionmatrix}) to the 
following form

\begin{eqnarray}
\label{collisionint}
\nonumber (\chi_{ij}, C^{2 \rightarrow 2} \chi_{ij}) &=& \frac{\beta^3}{(4\pi)^6} \sum_{\text{abcd}} 
\int_0^{\infty} q^2 dq ~ p^2 dp ~ k^2 dk \int_{-1}^{1} d \ \text{cos} \theta_{pq} ~ d \ \text{cos} 
\theta_{kq} \int_0^{2\pi} d\phi \ \frac{|\mathcal{M}^{ab}_{cd}|^2}{E_p E_k E_{p^\prime} E_{k^\prime}} 
\times \\
\nonumber && \delta(E_p+E_k-E_{p^\prime}-E_{k^\prime}) ~ f_{eq}^a(E_p) ~ f_{eq}^b(E_k) ~ 
[1-f_{eq}^c(E_{p^\prime})] ~ [1-f_{eq}^d(E_{k^\prime})] \times \\
&& \left[\chi_{ij}^a({\mathbf p})+\chi_{ij}^b({\mathbf k})-\chi_{ij}^c({\mathbf p^\prime})-
\chi_{ij}^d({\mathbf k^\prime}) \right]^2
\end{eqnarray}

At this point a new integration variable $\omega$, which corresponds physically to the energy 
transferred in the collision, is introduced following Baym et al.~\cite{leadinglogbaym} to do the angular 
integrals as follows.

\begin{equation}
\delta\left(E_p+E_k-E_{p^\prime}-E_{k^\prime}\right) = \int_{-\infty}^{+\infty} d\omega ~ 
\delta\left(\omega+E_p-E_{p^\prime}\right) \delta\left(w-E_k+E_{k^\prime}\right)
\end{equation}
Using the properties of the delta function one finds that the above delta functions can be written 
instead as

\begin{eqnarray}
\nonumber \delta \left(\omega+E_p-E_{p^\prime}\right) &=& \frac{E_{p^\prime}}{pq} 
\delta\left(\text{cos}\theta_{pq}-\frac{w}{q}\left(\frac{E_p}{p}\right)-\frac{t}{2pq} \right) 
\Theta\left(w+E_p\right) \\
\delta\left(w-E_k+E_{k^\prime}\right) &=& \frac{E_{k^\prime}}{kq} 
\delta\left(\text{cos}\theta_{kq}-\frac{w}{q}\left(\frac{E_k}{k}\right)+\frac{t}{2kq} \right) 
\Theta\left(E_k-w\right) 
\end{eqnarray}
with $\Theta$ denoting the unit step function. Carrying out these delta function integrals reduces 
Eq.~(\ref{collisionint}) further to the form below

\begin{eqnarray}
\label{simplifiedcollision}
\nonumber (\chi_{ij}, C^{2 \rightarrow 2} \chi_{ij}) &=& \frac{\beta^3}{(4\pi)^6} \sum_{\text{abcd}}
\int_0^\infty dq \int_{w_-}^{w_+} dw 
\int_{p_-}^\infty dp 
\int_{k_-}^\infty dk \int_0^{2\pi} d\phi ~ |\mathcal{M}^{ab}_{cd}|^2 ~ \left(\frac{p k}{E_p E_k}\right) 
\times \\ \nonumber && f_{eq}^a(E_p) ~ f_{eq}^b(E_k) ~ [1-f_{eq}^c(E_{p^\prime})] ~ 
[1-f_{eq}^d(E_{k^\prime})] \times \\
&& \left[\chi_{ij}^a({\mathbf p})+\chi_{ij}^b({\mathbf k})-\chi_{ij}^c({\mathbf p^\prime})-
\chi_{ij}^d({\mathbf k^\prime}) \right]^2
\end{eqnarray}
where the upper and lower limits of the $\omega$ integral are 
$\omega_{\pm}=\pm\left({\sqrt{q^2+M^2}-M}\right)$, 
the lower limit of the $p$ integral is $p_-=\left(q-w\sqrt{1-4M^2/t}\right)/2$, and that of the $k$ 
integral is $k_-=\left(q+w\sqrt{1-4M^2/t}\right)/2$. In the above equation, $E_{p^\prime}=E_p+w$ and 
$E_{k^\prime}=E_k-w$. The Mandelstam variable $t$ is as usual $t=w^2-q^2$. Imposing a 3-momentum cutoff in the Boltzmann equation is implemented by setting the upper limit 
on the q integral in Eq.~(\ref{simplifiedcollision}) above to twice the value of the 3-momentum cutoff $\Lambda$. 

In the t-channel parametrization, the other Mandelstam variables $s$ and $u$ are 
related to $t$ as given below.
\begin{eqnarray}
\nonumber s&=&2M^2-\frac{t}{2q^2} \left[ (E_p+E_{p^\prime})(E_k+E_{k^\prime})+q^2 \right]-\frac{\text{cos} \phi}{2q^2} \\
&& \times \sqrt{ \left[ (-t)(4E_p E_{p^\prime}+t)-4q^2M^2 \right] \left[  (-t)(4E_k E_{k^\prime}+t)-4q^2M^2 \right]} \ , \\
u &=& 4M^2-s-t \ ,
\end{eqnarray}
where $M$ is the constituent mass of the u and d quarks. Finally, one needs the cosine of the angles 
between various momenta and these are given below.

\begin{eqnarray}
\text{cos} \theta _{pq} = \left( \frac{E_p}{p} \right) \frac{w}{q}+ \frac{t}{2pq} \ , && \ \text{cos} \theta _{kq} = \left( \frac{E_k}{k} \right) \frac{w}{q}- 
\frac{t}{2kq} \ , \\
\text{cos} \theta _{p p^\prime} = \frac{E_p E_{p^\prime}-m^2}{p p^\prime}+ \frac{t}{2p p^\prime} \ , && \ \text{cos} \theta _{k k^\prime} = 
\frac{E_k E_{k^\prime}-m^2}{k k^\prime}+ \frac{t}{2k k^\prime} \ , \\
\text{cos} \theta _{p k^\prime} = \frac{E_p E_{k^\prime}-m^2}{p k^\prime}+ \frac{u}{2p k^\prime}\ , && \ \text{cos} \theta _{p^\prime k} = 
\frac{E_{p^\prime} E_k-m^2}{p^\prime k}+ \frac{u}{2p^\prime k} \ .
\end{eqnarray}
The neccessary equations for the chirally symmetric phase in which the constituent mass of quarks vanishes can be obtained from those above by simply setting $M=0$. 
There are equations similar to the ones above for the s-channel parametrization, however those will not be reproduced here.

\end{document}